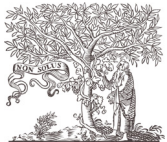
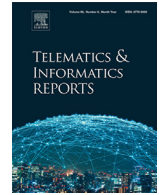

# Navigating the acceptance of implementing business intelligence in organizations: A system dynamics approach

Mehrdad Maghsoudi [a],[*], Navid Nezafati [b]

[a] *Department of Industrial and Information Management, Faculty of Management and Accounting, Shahid Beheshti University, Tehran, Iran*
[b] *BPP University, Business School, London, England*

ARTICLE INFO

*Keywords:*
Business intelligence
System dynamics
Implementation of business intelligence
Information technology organizations
Vansim

ABSTRACT

The rise of information technology has transformed the business landscape, with organizations increasingly relying on information systems to collect and store vast amounts of data. To stay competitive, businesses must harness this data to make informed decisions that optimize their actions in response to the market. Business intelligence (BI) is an approach that enables organizations to leverage data-driven insights for better decision-making, but implementing BI comes with its own set of challenges. Accordingly, understanding the key factors that contribute to successful implementation is crucial.

This study examines the factors affecting the implementation of BI projects by analyzing the interactions between these factors using system dynamics modeling. The research draws on interviews with five BI experts and a review of the background literature to identify effective implementation strategies. Specifically, the study compares traditional and self-service implementation approaches and simulates their respective impacts on organizational acceptance of BI. The results show that the two approaches were equally effective in generating organizational acceptance until the twenty-fifth month of implementation, after which the self-service strategy generated significantly higher levels of acceptance than the traditional strategy. In fact, after 60 months, the self-service approach was associated with a 30% increase in organizational acceptance over the traditional approach. The paper also provides recommendations for increasing the acceptance of BI in both implementation strategies. Overall, this study underscores the importance of identifying and addressing key factors that impact BI implementation success, offering practical guidance to organizations seeking to leverage the power of BI in today's competitive business environment.

## Introduction

The speed of data generation and accumulation has increased with the increasing use of information technology solutions by organizations, which are now using digital tools to store and analyze vast amounts of data in real-time [28,37,63]. Companies and organizations can gain a comparative advantage and overtake their competitors through the analysis and continuous and effective use of data and information [12,30,68].

Business intelligence is one of the popular and welcomed solutions for organizations to use data analysis for decision-making and data-oriented business [67,74]. For example, a retail company can use business intelligence to monitor customer behavior and preferences and adjust its marketing strategy accordingly [48]. Likewise, a manufacturing company can use BI to optimize its supply chain, reduce production costs, and improve product quality [27]. This solution offers managers and experts of organizations the possibility of making smart and updating analyzes and decisions. A business intel-ligence system is usually defined as a set of technological solutions [12] that facilitate organizations to collect, integrate, and analyze large data stores in order to understand their opportunities, strengths, and weaknesses [23].

According to researchers [1], business intelligence systems are quite close to the concept of decision support systems, these systems expand the categories of users and support a wide range of decisions. Business intelligence systems are designed to reduce uncertainty in the decision-making process and support decision-makers efficiently and effectively [51].

The size of the business intelligence market in 2020 was valued at $23.1 billion and is expected to reach $33.3 billion by 2025 and experience a compound annual growth rate of 6.6% [70]. However, despite growing investments and market expansion, evidence shows that many organizations cannot take advantage of the benefits of implemented business intelligence systems [5]. More than 70% of business intelligence projects cannot bring the expected returns [1] or result in little or no benefits for organizations [72]. To find the best way to use the value of business intelligence systems and succeed in their implementation [69].

Individual and organizational acceptance is one of the main challenges in the successful implementation of business intelligence systems in organizations, which is very complex due to their human nature and requires careful monitoring and control [65]. The acceptance of users is crucial for the successful utilization of business intelligence (BI) systems over an extended period. When users embrace BI and use it consistently, it becomes compatible with other organizational processes. Besides, BI can facilitate organizational change that leads to improvements in coordination and control processes. User acceptance is fundamental to implementing information system projects as a whole [14,24]. Specifically, user acceptance is critical when it comes to BI systems [15,21,32,55,61,71]. Many researchers have talked about the importance of organizational acceptance in the successful implementation of the business intelligence system in the organization. For example, [6] have used the Technology Acceptance Model (TAM) method to accept business intelligence, [22] have used the exploratory approach to conceptualize the acceptance of business intelligence, [39] has used motivation theory to analyze the two modes of routine use and innovative use. have benefited [21] is also a mixed study with the aim of investigating the factors that affect the acceptance behavior in the field of business intelligence. However, it is worth noting that, despite the extensive research on organizational acceptance of business intelligence systems, so far no article has used system dynamics to analyze this phenomenon. Therefore, this article aims to fill this research gap by using a dynamic system method to analyze the organization's acceptance of BI systems.

System dynamics is a powerful approach that can help researchers gain a deeper understanding of the complex interrelationships between various factors influencing organizational acceptance. System dynamics is an approach that enables researchers to model the complex interactions between different factors [38] influencing organizational acceptance. By developing simulation models that capture the dynamic behavior of the system over time, researchers can identify the root causes of problems and test different policies and strategies to improve user acceptance. By employing a system dynamics approach, researchers can develop simulation models that capture the dynamic behavior of the system over time, which can provide valuable insights into the long-term consequences of different policies and strategies. Therefore, there is a need for more research that utilizes system dynamics to investigate organizational adoption of business intelligence systems [52]. After stating the generalities of the research and the statistics of the factors affecting the successful implementation of business intelligence projects, this research has drawn the cause-effect relationships of these factors based on the opinions of experts and simulated the dynamic behavior of organizational acceptance in two traditional and self-service strategies and based on The results of that analysis provide.

This article is divided into seven sections. In Section 2, we provide an overview of existing literature on business intelligence and system dynamics, including strategies for implementing business intelligence systems, and review associated works. Section 3 discusses the research methodology used in this study. In Section 4, we present our findings on the identification of factors that affect the implementation of business intelligence systems. We also showcase the results of our dynamic system modeling and simulation of different scenarios. Section 5 is dedicated to the Conclusion and Outlook of the Research. Section 6 is dedicated to discussing the limitations of our study. While we have made every effort to conduct a comprehensive study, it is possible that some factors were not accounted for due to these limitations. Finally, in Section 7, we outline future research directions that could build upon our work.

**Research literature review**

*Business intelligence*

The term business intelligence was first proposed in 1989 by the Gartner Group. They introduced business intelligence as a set of concepts and methods to develop business decisions through reality-based systems [57]. Business intelligence includes all the processes of collecting, storing, accessing, analyzing, and extracting quality information or knowledge in different business fields [50]. A business intelligence system enables employees and organizations to better understand their business or market and make timely strategic decisions [47]. Business intelligence is a concept that has evolved over time. In the beginning, business intelligence was mainly focused on data analysis, but today it also includes organizational processes and strategies because business intelligence affects not only technology but also the organization that applies business intelligence [54]. Business intelligence has also been developed and used at different levels of the organization, from the strategic level to the operational level, in order to make more decisions based on data [2].

The importance of business intelligence lies in its ability to provide organizations with valuable insights that can enhance their competitiveness and profitability [47]. BI can help organizations improve their operational efficiency, reduce costs, increase revenue, and identify new business opportunities [46]. By having access to timely and accurate data, organizations can make better decisions, mitigate risks, and respond to changing market conditions more effectively. In today's fast-paced and data-driven business environment, companies that fail to leverage the power of business intelligence risk falling behind their competitors and missing out on growth opportunities [10].

*System dynamics*

System dynamics appeared in the late 1950s as a result of focusing on the behavior of complex systems in a specific period [29]. The main features of system dynamics simulation are feedback loops, state-flow functions, and time delays which are used to model the nonlinearity of systems' behavior [60].

In system dynamics, when a set of variables are connected to each other in a connected path, they form a feedback loop, which includes positive feedback loops and negative feedback loops. Positive feedback loops are circles in which if a factor is changed in one direction, the circle reinforces the changes in that direction. Negative feedback loops are circles that, if a factor is changed in one direction, the circle opposes

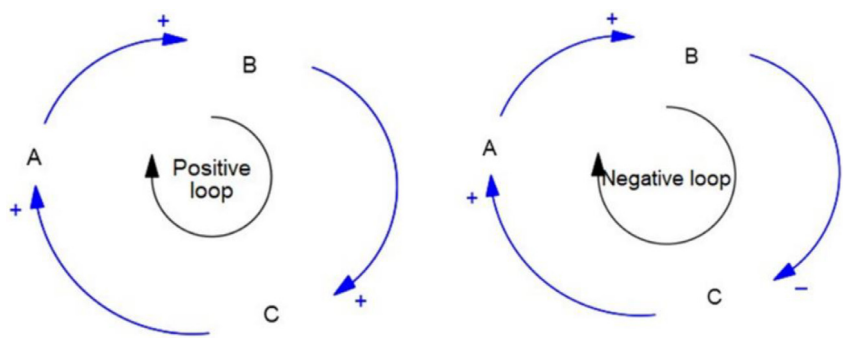

**Fig. 1.** Positive feedback loop (left) and negative feedback loop (right) [9].

changes in the factor in that direction [17]. Fig. 1 shows positive and negative feedback loops.

In system dynamics, dynamic variables move through currents in the system and accumulate in the state variable. A state variable calculates and reports the quantitative state of dynamic variables at any given time [25]. Feedback loops and delays in system dynamics can simulate a dynamic system, that is, if a change occurs at any point in the system, it will lead to a chain reaction throughout the system [53]. System dynamics allows for an iterative process of model-building, testing, and refinement to gain insights into the behavior of the system over time [7]. The approach is widely used in various fields such as engineering, economics, and management to understand and optimize the performance of complex systems.

Compared to other methods such as Delphi, system dynamics has several advantages. Firstly, it allows for the explicit modeling of feedback loops, which are often present in complex systems but difficult to capture using traditional modeling approaches. Secondly, system dynamics incorporates the concept of time delays, which can have significant impacts on the behavior of the system over time. Thirdly, the iterative nature of system dynamics modeling allows for continuous improvement and refinement of the model based on new data or insights. Finally, system dynamics provides a visual representation of the system, which can facilitate communication and understanding among stakeholders with varying levels of expertise.

The general process of system dynamics involves the following steps [41,43,56]:

1. Identifying the problem: The first step in system dynamics is to identify the problem or issue that needs to be analyzed. This could be something like a declining sales trend in a business or an increase in unemployment in a community.
2. Building a conceptual model: Once the problem has been identified, a conceptual model is built to represent the system being analyzed. This model includes all the relevant variables, relationships between variables, and feedback loops.
3. Quantifying the model: The next step is to quantify the conceptual model by assigning numerical values to the variables and relationships in the model. This allows for mathematical analysis of the system and prediction of how it will behave over time.
4. Simulation and testing: After the model has been quantified, it is simulated using computer software to test its behavior under different conditions. This helps to identify the causes of the problem and potential solutions.
5. Policy design and implementation: Based on the results of the simulation, policy recommendations are developed to address the problem. These policies are implemented, and their impact is monitored over time to assess their effectiveness.

*Types of strategies in implementing business intelligence*

In general, there are two methods and strategies for deploying business intelligence in organizations. Implementation of business intelligence in the traditional way and self-service business intelligence [2,34,35], each of these two methods has its own characteristics:

*Traditional BI implementation*

Traditional BI implementations follow a centralized approach, where IT departments take complete control over the BI system development process [26]. This implementation method involves various stages such as data warehousing, ETL (Extract, Transform, Load), data modeling, report generation, analysis, and distribution. These stages require specialized skills and expertise, which are generally available in the IT department [3]. The following are some of the critical elements of traditional BI implementation:

· **Data Model Design**: Traditional BI implementation relies heavily on the IT team to develop the data model. This is done by mapping the business requirements to the data warehouse schema. The IT team uses tools like ER diagrams and dimensional modeling techniques to create the data model(David [13]).
· **Data Extraction, Transformation, and Loading**: Data extraction involves collecting data from various sources and transforming them into a standard format. The transformed data is then loaded into the data warehouse. This process requires significant effort and expertise [40].
· **Report Generation**: Once the data is ready, reports are generated using specialized tools like Crystal Reports, Business Objects, or Cognos [75]. The IT team creates custom reports based on the business requirements, which can be published and distributed to end-users [33].
· **Analysis and Distribution**: Traditional BI implementation involves the creation of static reports that are distributed to end-users for analysis. End-users have limited options to manipulate the data, and any changes require IT involvement [20].

*Self-service BI implementation*

Self-Service BI implementation follows a decentralized approach, where end-users have more control over the BI system development process. Self-service BI provides end-users with more flexibility and agility in accessing and analyzing data [36,49]. The following are some of the critical elements of self-service BI implementation:

· **Data Model Design**: Self-service BI implementation allows end-users to create their own data models using simple drag-and-drop interfaces. End-users can create data models based on their business requirements without any IT involvement [49].
· **Data Extraction, Transformation, and Loading**: Self-service BI implementation allows end-users to collect data from various sources and transform them into a standard format. This process does not require any specialized skills or knowledge [4].
· **Report Generation**: Self-service BI implementation provides end-users with tools like Tableau, QlikView, or PowerBI to create custom reports. These tools have user-friendly interfaces that allow end-users to create reports based on their specific needs [16].
· **Analysis and Distribution**: Self-service BI implementation enables end-users to analyze and manipulate data on their own. They can create custom dashboards, drill-downs, and filters to explore data in real time. End-users can share their findings with others using interactive reports that can be accessed on any device [59].

Table 1 summarizes the differences between traditional and self-service BI implementation methods:

**Table 1**
Differences between traditional and self-service BI.

| Task | Traditional BI Implementation | Self-Service BI Implementation |
|---|---|---|
| **Data Model** | Developed by IT team | Developed by End-users |
| **ETL** | Done by IT team | Done by End-users |
| **Report Gen.** | Done by IT team | Done by End-users |
| **Analysis** | Limited options for end-users | Flexible options for end-users |
| **Control** | Centralized | Decentralized |
| **Expertise** | Requires specialized skills | No specialized skills required |

*Associated works*

Nalchiger et al., in a research they conducted in 2014, presented an approach based on systems dynamics modeling to support decisions and choose the best alternative actions for the organization through the outputs of the business intelligence system. These researchers modeled the results of the business intelligence system with the decision-making process in the organization and actually presented a combination of business intelligence and a decision support system [45].

Ain et al. also investigated the reason why organizations do not achieve the benefits of implementing a business intelligence system and using a systematic review of past research, provided comprehensive knowledge about what has been stated in the field of acceptance, use, and success of a business intelligence system. They say These researchers reviewed 11 related articles and identified the effective factors related to the adoption, use, and success of business intelligence [1].

**Table 2**
Summary of related works.

| Year | Authors | Subject | Key Findings |
| --- | --- | --- | --- |
| 2014 | Nalchiger et al. | Systems Dynamics Modeling and Business Intelligence for Decision Support | Systems dynamics modeling can support decision-making in organizations through business intelligence system outputs. Combination of BI and decision support system presented. |
| 2019 | Ain et al. | Factors Affecting Adoption, Use and Success of Business Intelligence Systems | Effective factors related to the adoption, use and success of business intelligence identified through a systematic review of past research. |
| 2020 | Kumar & Krishnamoorthy | Adoption of Business Analysis Systems in India | Data quality and human resource competencies are main challenges; technology assets and competitive pressure are driving factors for organizations to adopt business analytics systems. |
| 2020 | Müller et al. | Success Factors of Implementing Business Intelligence in Medium and Large Organizations | Top management support, information technology infrastructure and system quality are of highest importance for success of business intelligence systems in medium and large organizations in food industry based on data analysis using structural equation modeling. |
| 2021 | Mehri | Critical Success Factors of Business Intelligence Projects in Public Sector | Information systems and data quality are most important factors among fourteen critical success factors of business intelligence projects in public sector categorized into organization, process and technology using hierarchical analysis. |
| 2022 | Fu et al. | Important Factors Considered by Companies to Introduce a Business Intelligence System | Company information is most important factor when introducing a business intelligence system in companies followed by system performance integrity, closeness to company's strategy, license cost and technology maturity after analyzing vital factors using fuzzy hierarchical analysis and VIKOR techniques. |

Kumar and Krishnamoorthy have made the adoption of business analysis systems the subject of their research and have studied the technological capabilities and adoption of business analysis systems in India. These two researchers obtained their data through semi-structured interviews and came to the conclusion that data quality and human resource competencies are the main challenges to the adoption of business analytics systems in India, and technology assets and competitive pressure are the main driving factors for organizations to pay attention to business analytics systems. are business analysis [31].

In another study, researchers have investigated the success factors of implementing business intelligence systems in medium and large organizations. In this research, the researchers went to the food industry and conducted their studies based on the data of 69 companies active in this industry. These researchers used structural equation modeling for data analysis based on which top management support, information technology infrastructure, and system quality are of the highest importance for the success of business intelligence systems in this industry [44].

Mehri has also evaluated the critical success factors of business intelligence projects in the public sector using hierarchical analysis. In this article, he identified fourteen main factors and after dividing them into three categories of organization, process, and technology, he prioritized the factors using the opinions of nine experts in this field. According to the findings of this project management researcher, information systems and data quality are the most important factors among the fourteen critical success factors [42].

In their research, Fu et al. investigated the important factors considered by companies to introduce a business intelligence system. This study has collected and analyzed the critical factors considered by companies when introducing a business intelligence system. By studying the research literature, these researchers have calculated all the vital factors before, during, and after the introduction of the business intelligence system and by using the two techniques of fuzzy hierarchical analysis and VIKOR, four factors of system performance integrity, closeness to the company's strategy, license cost, and technology maturity. Company information has been selected as the most important factor when introducing a business intelligence system in companies [18]. Overall, these studies reveal that the successful adoption and implementation of BI systems depend on various factors such as organizational support, technological capabilities, data quality, and strategic alignment. The findings of these studies can guide organizations in making informed decisions about implementing BI systems to enhance decision-making processes and improve overall business performance. Table 2 provides a summary of related works:

**Methodology**

In terms of both purpose and methodology, this study is pragmatic and falls under the category of survey research. Fig. 2 illustrates the approach taken in conducting the study. The simulation was conducted at a granular level on the surface of the moon with a time horizon of 5 years. This timeframe was determined based on expert opinion, as various factors within organizational environments often impact one another with some lag time, resulting in delayed effects. System dynamics modeling allows for incorporating such delays into the analysis [73].

In this study, we undertake a comprehensive process to identify the factors that influence the implementation of business intelligence systems. This involves an extensive review of the subject literature and previous studies to compile a list of potential factors. We present the extracted factors to a panel of five experts who provide feedback and confirmation on their relevance. To ensure the validity of the identified factors, the selection criteria require approval from a minimum of three out of the five experts present.

Following the determination of effective factors, close collaboration between the research team and experts leads to the establishment of relationships between these factors, as well as the initial values associated with the exogenous factors for each of the implementation strategies. We use this information to create a system dynamics model, which we evaluate and simulate in both traditional and self-service strategies. In the final stage of the research, we compare and analyze the simulation results from both strategies.

Choosing a panel of five experts to provide feedback and confirmation on the relevance of extracted factors is necessary for this study for several reasons. Firstly, it ensures that multiple perspectives are considered when selecting the factors that influence the implementation of business intelligence systems. With more people involved in the selection process, there is a greater chance of identifying factors that may have been overlooked by one person. Secondly, requiring approval from a minimum of three out of the five experts present ensures the validity of identified factors. This criterion provides a higher level of confidence in the selected factors, as they have been evaluated and approved by a majority of experts. Lastly, close collaboration between the research team and the chosen experts enables the establishment of relationships between the identified factors and the initial values associated with exogenous factors for each implementation strategy. This collaboration facilitates a better understanding and interpretation of the data and results. In turn, this improves the accuracy of the system dynamics model created, which is essential for evaluating and simulating both traditional and self-service strategies.

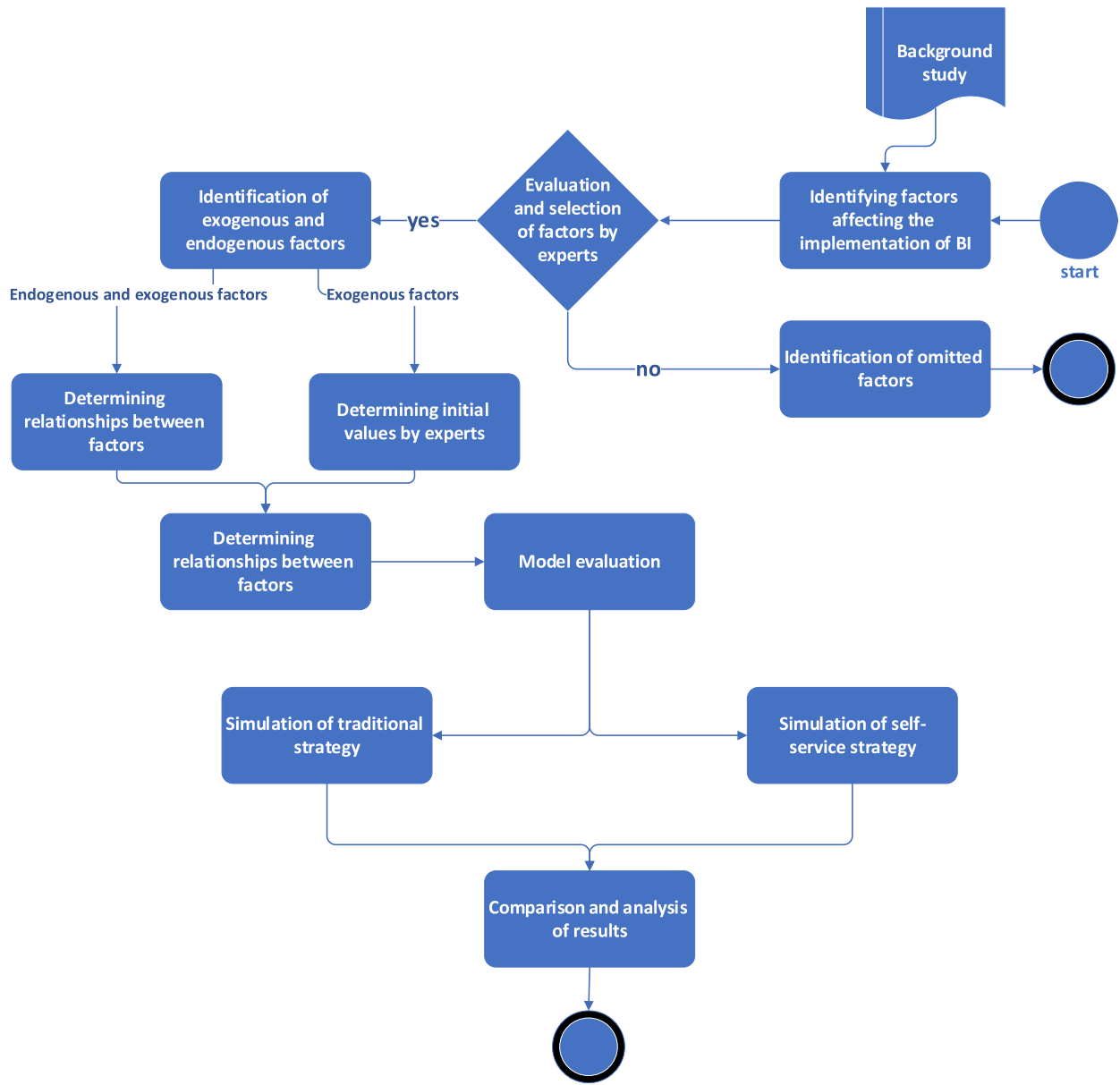

**Fig. 2.** Research model.

**Table 3**
Profile of experts.

| gender | position | education | work experience | Organization size |
| --- | --- | --- | --- | --- |
| Female | Head of BI Department | Masters | 16 years | More than 2000 |
| Man | Data engineering team manager | P.H.D | 6 years | More than 400 |
| Man | BI project manager | Masters | 5 years | More than 500 |
| Man | technical manager | Masters | 18 years | More than 200 |
| Female | BI specialist | PhD candidate | 5 years | Different |

It is worth noting that additional information about the participating experts can be found in Table 3, providing further insight into their contributions towards the identification of the factors affecting the implementation of business intelligence systems.

**Result**

*Identifying factors affecting the implementation of BI*

Many researchers [1,14,19] have tried to identify factors affecting the implementation of business intelligence systems.

The set of indicators available in the literature is described in Table 4:

*Evaluation and selection of factors by experts*

Since the high number of factors affecting the implementation of business intelligence, which leads to the high complexity of the dynamic model, the list of factors in the form of a questionnaire was presented to five research experts to select the main factors affecting the implementation of the business intelligence system, and it was established that if at least three The effect of the factor has been confirmed and that factor

**Table 4**
List of factors affecting the implementation of business intelligence.

| Factors | References | Factors | References | Factors | References | Factors | References |
|---|---|---|---|---|---|---|---|
| Project Management Skills | 1, 2, 3, 4 | Domain knowledge of BI team | 2, 3 | Project leadership power | 4 | Service quality | 1, 2, 4 |
| Net Benefits | 1, 2, 3, 4 | Ability to manage change | 1, 2, 3, 4 | External environment | 4 | Ease of use of the system | 1 |
| User participation | 1, 2, 3, 4 | Communication strength of the BI team | 2,3 | Technical readiness of BI | 4 | User satisfaction | 2, 3, 4 |
| Timely response | 1 | System integrity | 1, 2, 3, 4 | Third-Party Interactions | 3,4 | Intention To Use | 1 |
| Data quality | 1, 4 | Hope for performance | 3, 4 | Developer Skills | 4 | Technology Experience | 1 |
| Ease of use | 1, 2, 4 | Organizational Learning | 2, 4 | Development Approach | 2, 4 | Attitude Towards Change | 1 |
| Training users | 2, 4 | BI system maturity | 2, 3, 4 | Organizational Structure | 2,4 | Trust | 2, 4 |
| Top management commitment | 1, 4 | Data maturity of the organization | 3,4 | Organizational Competence | 2,4 | User Expectations | 2, 4 |
| IT Infrastructure | 1, 2, 4 | Manager's social influence | 1,2 | Organizational Size | 4 | Subjective Norms | 2, 3, 4 |
| System quality | 1, 2 | Organization acceptance | 2,3,4 | Organizational Culture | 1,2,4 | Teamwork & composition | 2, 4 |
| Access to organization data | 1 | Definition of clear vision | 2,3 | Competency Development | 4 | Management support | 1, 2, 3, 4 |
| Ability to integrate with other systems | 1 | Fear of losing organizational status | 2, 3, 4 | Human Resources | 1 | Knowledge and technical capabilities of the BI team | 2, 4 |

**1:** [19], **2:** [1], **3:** [14], **4:** [65]

**Table 5**
Factors selected by experts.

| | |
|---|---|
| Definition of clear vision | IT Infrastructure |
| Top management commitment | Ability to integrate with other systems |
| Ease of use of the system | Hope for performance |
| User participation | Organizational Learning |
| Training users | BI system maturity |
| Data quality | Data maturity of the organization |
| Access to organization data | Project management ability |
| System quality | Ability to manage change |
| Timely response | Communication strength of the BI team |
| System integrity | Organization acceptance |
| Management support | Domain knowledge of BI team |
| Knowledge and technical capabilities of the BI team | Fear of losing organizational status |
| | Manager's Social influence |
| | Organization size |

**Table 6**
boundary of the model.

| Endogenous variable | Exogenous variable | Omitted variables |
|---|---|---|
| User participation | Knowledge and technical capabilities of the BI team | Organization size |
| Organization acceptance | | Organizational Learning |
| Data quality | Domain knowledge of BI team | Hope for performance |
| Ease of use of the system | | Ability to integrate with other systems |
| Access to organization data | Communication strength of the BI team | BI system maturity |
| IT Infrastructure | | Ability to manage change |
| System quality | Fear of losing organizational status | Project management ability |
| Training users | | Data maturity of the organization |
| Timely response | | Manager's Social influence |
| System integrity | | |
| Management support | | |
| Top management commitment | | |
| Definition of clear vision | | |

should be presented in the model. The result of the selection of experts is described in Table 5:

*Identification of exogenous and endogenous factors*

In the field of economics and social sciences, it is essential to understand the dynamics of a system and how different variables interact with each other [64]. In this context, three categories of variables are often used to analyze and model complex systems: endogenous, exogenous, and omitted variables [58,62].

- Endogenous variables are those that are within the system being analyzed and are influenced by other variables within that same system. These variables are typically the focus of the analysis, as they represent the dependent or outcome variable(s) of interest. These variables are important because they help us understand how changes in one part of the system can affect other parts of the system [43].
- Exogenous variables, on the other hand, are external to the system being analyzed and are not influenced by other variables within that same system. Instead, these variables are typically considered to be independent or causal factors that affect the behavior of the endogenous variables. These variables are important because they help us understand the broader context in which the system operates and how it might be affected by external forces [66].
- Finally, there are omitted variables, which are simply those that are not included in the analysis of the system. These variables may be relevant to the behavior of the system, but for various reasons (such as lack of data or the complexity of the system) they are not considered in the analysis [62].

Complex dynamic systems can be effectively analyzed and predicted by understanding the role of these different types of variables [11]. By carefully considering the relationships between endogenous and exogenous variables, and striving to identify any potentially relevant omitted variables, analysts can gain a deeper understanding of how the system works and make more accurate predictions about its future behavior [43].

Table 6 shows the boundary diagram of the research. Endogenous factors are factors whose behavior changes in the model according to other variables. Exogenous factors are factors that are not affected by other variables and their behavior is not affected by the model. The reason for leaving out some factors is to prevent the enlargement of the model and based on the opinions of experts.

*Causal loop diagram*

System dynamics models have a crucial feature in feedback loops where positive feedback loops are referred to as reinforcing, symbolized by + or R, and negative feedback loops are called balancing, and represented by - or B. This is because positive loops amplify changes while negative loops self-correct.

Fig. 3's cause-effect diagram displays the connections between the factors listed in Table 6 and their respective positive or negative impacts on each other.

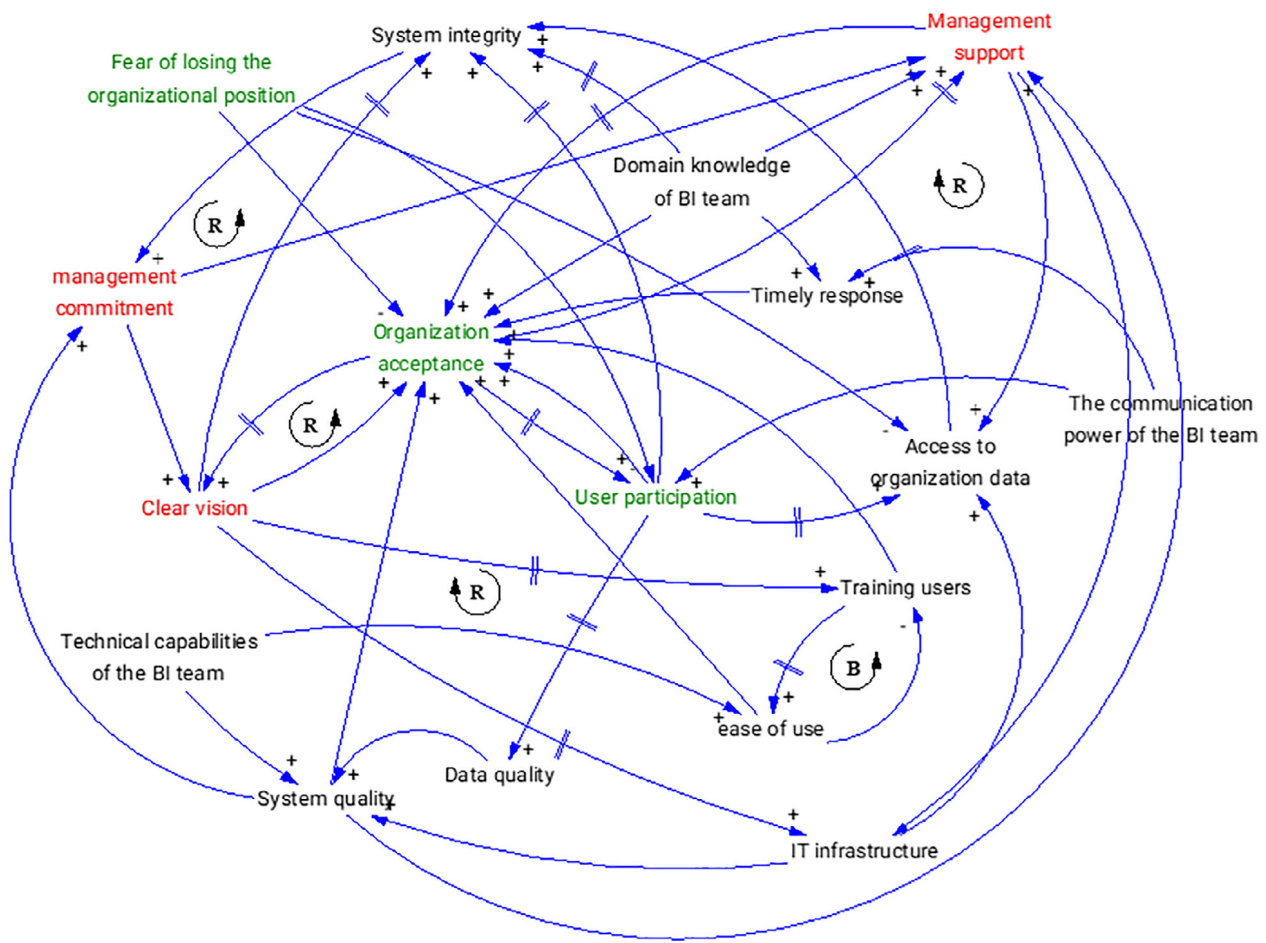

**Fig. 3.** Cause and effect diagram.

*Stock-flow diagram*

The flow diagram of the model highly relies on stock and flow variables.

Fig. 4 illustrates the prepared stock-flow diagram for the problem, based on the determined cause-and-effect diagram and five expert opinions. Stocks are depicted as rectangles while streams take on the form of arrows entering or leaving the stocks. In system dynamics modeling, stock and flow concepts are fundamental, with all system equations being written based on these concepts. Stocks denote the state of the system at a specific time and possess properties of aggregation [8]. Their values can increase or decrease through input and output currents, with the stock value obtained from the difference between the two. The remaining variables are covariates that impact not only each other (blue arrows) but also flows and stock.

Formulas were assigned to each variable using the problem flowchart. At this point, expert opinions served as the determining factor. In the initial stage, experts' views were consulted to determine the state variables' values and their rates, based on current Iranian organization conditions. Closed interviews were conducted to gather these opinions, and Table 7 displays the average results of these sessions.

*Validation of the model*

The model's validation is a prerequisite for utilizing and analyzing the results, as failure to verify its accuracy renders it unusable. In this research, validation is determined through a sentiment analysis test involving variable modifications. Specifically, Fig. 5 depicts changes in the "Fear of Losing Organizational Position" variable, revealing that or-ganizational acceptance decreases significantly when fear is quadrupled (blue line). This natural behavior in the model serves as evidence of its validity.

**Table 7**
Default conditions considered for exogenous variables based on experts' opinion.

| Variable name | The initial numerical value is considered according to the opinion of experts | |
|---|---|---|
| | Self-service | Traditional |
| **Technical knowledge of the BI team** | 0.8 | 1 |
| **Domain knowledge of the BI team** | 0.9 | 0.6 |
| **Communication strength of the BI team** | 0.8 | 0.4 |
| **Fear of losing position** | 0.3 | 0.8 |

*Results of model execution*

Once the model is validated through sensitivity testing, it becomes reliable for analysis following implementation. The system's dynamics are taken into account during the model's implementation, with a focus on the positive and negative interactions between all factors involved. Fig. 6 depicts the results of implementing the model in both traditional and self-service strategies. The model operates over a five-year time horizon and at the monthly level of granularity.

As it is clear from Fig. 6, the behavior of the organizational acceptance variable in traditional and self-service strategies is very close to each other and the direction of the numbers of this variable in each of the months is shown in Table 8.

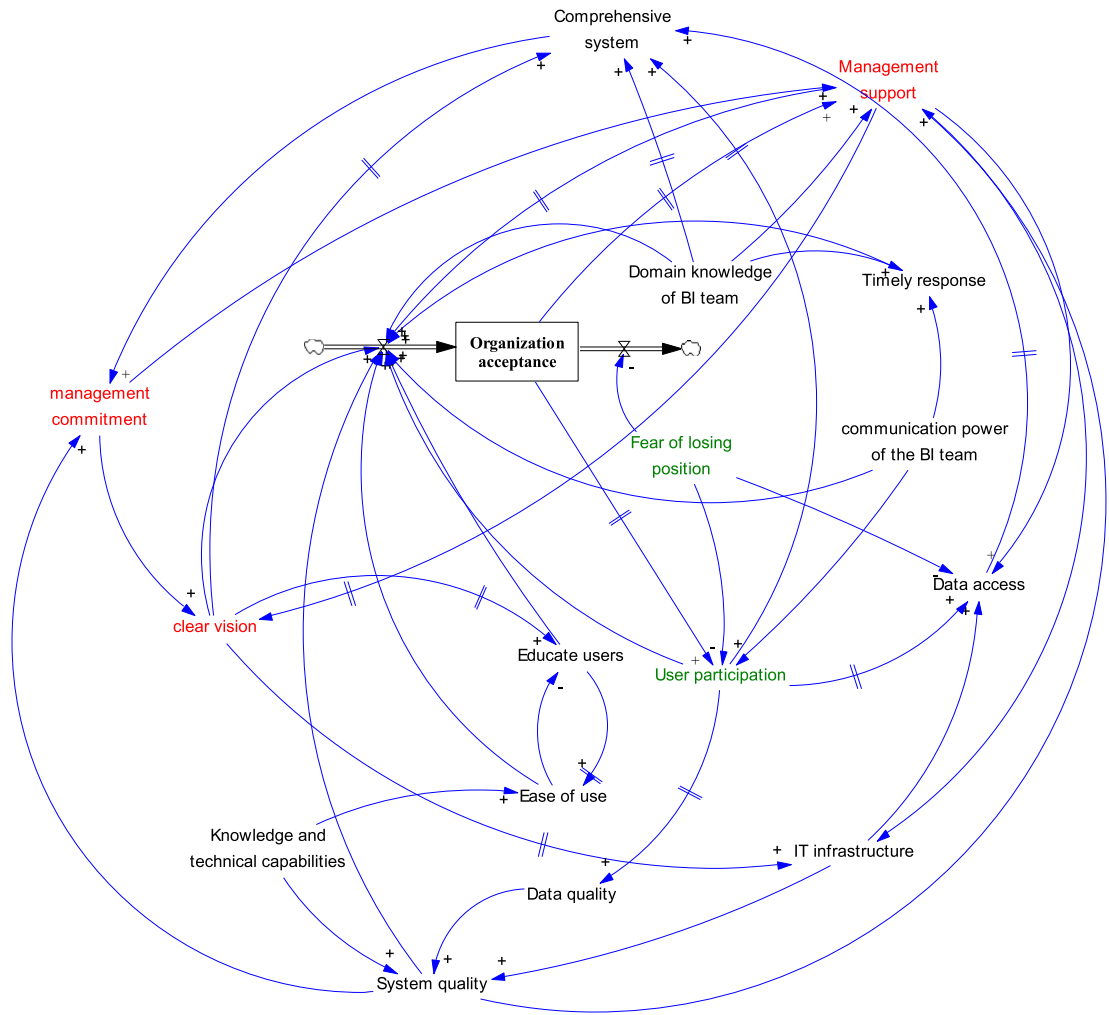

**Fig. 4.** Stock-flow diagram.

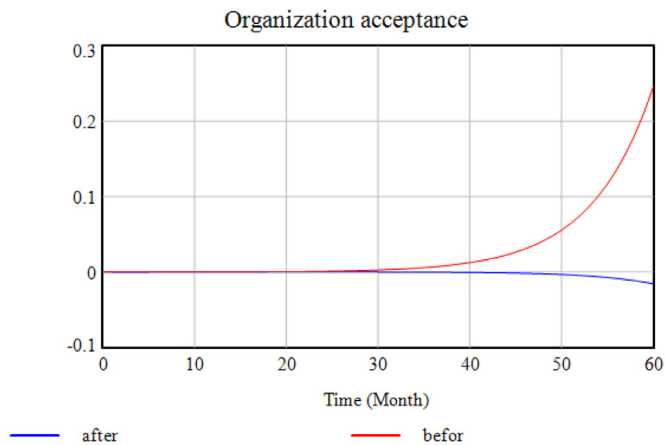

**Fig. 5.** Diagram of sensitivity analysis of organizational acceptance behavior with changes in the variable of fear of losing organizational position.

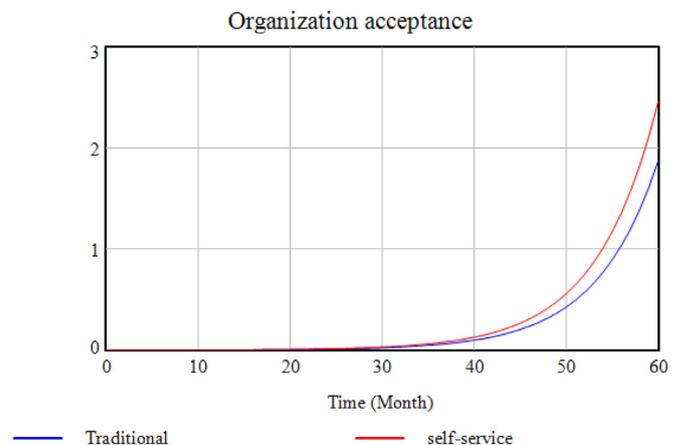

**Fig. 6.** The variable behavior of organizational acceptance in traditional strategies (red) and self-service (blue).

As it is clear from Fig. 5 and its values in Table 8., the amount of organizational acceptance variable was the same until the twenty-fifth month from the start of project implementation in both traditional and self-service strategies, but after that, the distance between these two strategies increased significantly. He found that the self-service strategy in the 60th month attracted 30% more organizational acceptance than the traditional strategy and is actually more successful.

*Check the scenarios*

Based on the initial results of the simulation of the two strategies, the self-service strategy has been more successful in attracting organizational acceptance and has overtaken the traditional strategy. However, many IT organizations use this strategy to implement and develop

**Table 8**
The amount of changes in organizational acceptance variable in traditional and self-service strategies.

| Month | Self-service | Traditional | Month | Self-service | Traditional | Month | Self-service | Traditional |
|---|---|---|---|---|---|---|---|---|
| 1 | 0.0000 | 0.0000 | 21 | 0.0060 | 0.0056 | 41 | 0.1460 | 0.1117 |
| 2 | 0.0001 | 0.0001 | 22 | 0.0071 | 0.0066 | 42 | 0.1694 | 0.1296 |
| 3 | 0.0001 | 0.0001 | 23 | 0.0083 | 0.0076 | 43 | 0.1966 | 0.1503 |
| 4 | 0.0002 | 0.0002 | 24 | 0.0097 | 0.0089 | 44 | 0.2282 | 0.1744 |
| 5 | 0.0003 | 0.0002 | 25 | 0.0113 | 0.0103 | 45 | 0.2648 | 0.2023 |
| 6 | 0.0004 | 0.0003 | 26 | 0.0132 | 0.0120 | 46 | 0.3073 | 0.2346 |
| 7 | 0.0005 | 0.0004 | 27 | 0.0153 | 0.0140 | 47 | 0.3566 | 0.2722 |
| 8 | 0.0007 | 0.0005 | 28 | 0.0179 | 0.0162 | 48 | 0.4138 | 0.3157 |
| 9 | 0.0009 | 0.0007 | 29 | 0.0208 | 0.0188 | 49 | 0.4802 | 0.3663 |
| 10 | 0.0011 | 0.0009 | 30 | 0.0242 | 0.0218 | 50 | 0.5572 | 0.4249 |
| 11 | 0.0013 | 0.0010 | 31 | 0.0281 | 0.0253 | 51 | 0.6465 | 0.4929 |
| 12 | 0.0016 | 0.0013 | 32 | 0.0327 | 0.0294 | 52 | 0.7501 | 0.5718 |
| 13 | 0.0019 | 0.0015 | 33 | 0.0380 | 0.0341 | 53 | 0.8704 | 0.6633 |
| 14 | 0.0022 | 0.0018 | 34 | 0.0442 | 0.0395 | 54 | 1.0099 | 0.7695 |
| 15 | 0.0027 | 0.0022 | 35 | 0.0513 | 0.0459 | 55 | 1.1717 | 0.8927 |
| 16 | 0.0032 | 0.0025 | 36 | 0.0596 | 0.0532 | 56 | 1.3595 | 1.0356 |
| 17 | 0.0037 | 0.0030 | 37 | 0.0692 | 0.0617 | 57 | 1.5773 | 1.2014 |
| 18 | 0.0044 | 0.0035 | 38 | 0.0804 | 0.0716 | 58 | 1.8301 | 1.3938 |
| 19 | 0.0052 | 0.0041 | 39 | 0.0933 | 0.0830 | 59 | 2.1233 | 1.6170 |
| 20 | 0.0060 | 0.0048 | 40 | 0.1083 | 0.0963 | 60 | 2.4635 | 1.8759 |

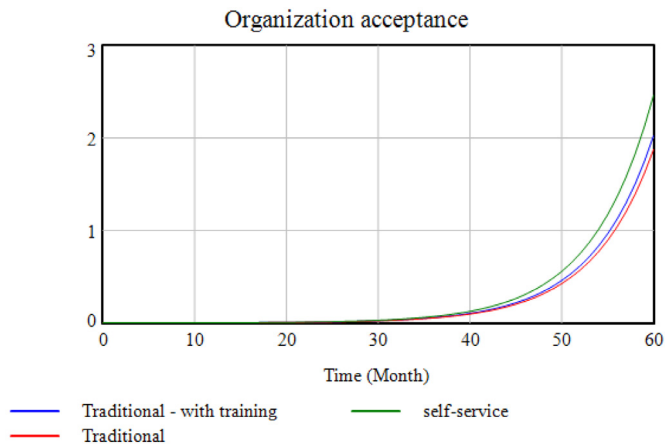

**Fig. 7.** Changing behavior of organizational acceptance with increasing domain knowledge.

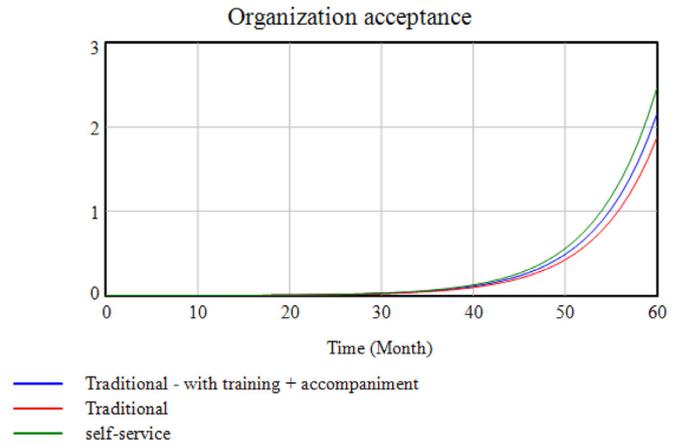

**Fig. 8.** Organizational acceptance variable behavior with increasing domain knowledge along with companionship.

their business intelligence system. For this reason, in this section, based on the opinions of experts, several scenarios for changes in the initial conditions of the model are discussed:

*First scenario: acquisition of domain knowledge*

According to the opinion of experts, one of the main shortcomings of the traditional method is the lack of sufficient Domain knowledge in the technical unit (business intelligence) regarding the activity of the unit requesting the dashboard and business intelligence system. To solve this shortcoming, experts suggest the solution of participating in training courses to acquire Domain knowledge of the system applicant. According to experts, this method increases people's knowledge by 20%. The results of this scenario are shown in Fig. 7.

Based on the results obtained from the implementation of the model in this scenario (blue line), the improvement of this situation compared to before after 60 months of project implementation is 11%, which makes the distance between traditional implementation and autonomous implementation reduced to 20%. And the traditional method will be more successful to gain organizational acceptance.

**The second scenario: acquiring domain knowledge along with increasing companionship**

Another scenario that was suggested by the experts is to increase cooperation with the requesting team (or the team for which the system is developed). One of the serious challenges in the implementation of the business intelligence system is the failure of experts to accompany it due to the fear of losing their position and organizational position. In this case, experts resist using the system by citing reasons such as business intelligence being useless, technical problems of having the system, inability to use the system, etc. Experts have recommended the solution of holding briefing meetings with the manager of the relevant team or the direct technical manager and have stated that this method will improve resistance by 25%. If these conditions are established together with the past scenario and the acquisition of domain knowledge of the team, the results of Fig. 8 will be created.

Based on the results of this scenario (blue line), the amount of gaining organizational acceptance in this scenario has improved by 19% compared to the initial state, and in this case, the difference between the traditional strategy and the self-service strategy in gaining organizational acceptance is 11%. has reduced.

**Third scenario: acquiring domain knowledge along with acquiring soft skills**

Another challenge mentioned by the experts, who have emphasized it, is the lack of communication skills in the dashboards development team, so these people have a major problem in fully delivering the system to the requester and also responding to their problems, which leads to challenges in the project. becomes Experts have stated the reason for this issue is the basic reliance of the business intelligence system development team on technical issues and lack of attention to soft skills. The solution provided by the experts to reduce this problem is holding soft skills and negotiation art courses for the development team. Experts have acknowledged that the effect of these courses (if held as a workshop and problem-oriented) improves conditions by 50%. In this scenario, the changes in the model will be as shown in Fig. 9.

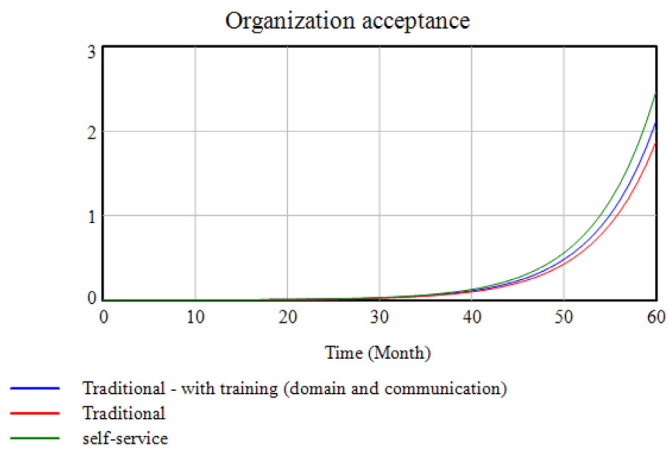

**Fig. 9.** Changing behavior of organizational acceptance with increasing domain knowledge and increasing communication skills.

In this scenario (blue line), organizational acceptance has experienced a 17% growth compared to the initial state of the traditional strategy, and this scenario reduces its distance from the self-service strategy to 13%.

*Discuss results*

The results of the model execution show that implementing the self-service strategy leads to higher levels of organizational acceptance compared to the traditional approach. The model operates over a five-year period at a monthly level of granularity, and it is evident from Fig. 6 and Table 8 that the organizational acceptance variable in both strategies was similar until the twenty-fifth month, after which the distance between the two increased significantly. Specifically, the self-service strategy attracted 30% more organizational acceptance than the traditional method and proved more successful overall.

However, to check the validity of the initial results, several scenarios were discussed based on expert opinions. The first scenario suggested acquiring domain knowledge, and the results showed an improvement of 11% in gaining organizational acceptance compared to the initial state, reducing the difference between the traditional and self-service strategies to 20%. The second scenario involved increasing cooperation with the requesting team, which improved resistance by 25%, and combining this with the acquisition of domain knowledge resulted in a further increase in gaining organizational acceptance by 19% compared to the initial state. The third scenario recommended acquiring soft skills through workshops and problem-oriented training, resulting in a 17% growth in organizational acceptance compared to the initial state of the traditional strategy, which reduced its difference from the self-service strategy to 13%.

Overall, these scenarios suggest that while the self-service strategy initially appeared more successful, there are ways to improve the traditional approach and make it more competitive. By addressing the lack of domain knowledge, increasing cooperation with the requesting team, and developing soft skills in the development team, organizations can increase their chances of successful implementation and gain organizational acceptance for their business intelligence systems.

**Conclusion and outlook of research**

The purpose of this research was to investigate the impact of business intelligence system implementation strategies in IT organizations on organizational acceptance. The study identified effective factors in the implementation of business intelligence systems through expert consultation and created a dynamic model to simulate the self-service and traditional approaches. The results showed that the self-service strategy led to higher levels of organizational acceptance compared to the traditional approach, with a difference of 30%.

To validate the initial results, three scenarios were proposed by experts: acquiring domain knowledge, increasing cooperation with the requesting team, and developing soft skills in the development team. These scenarios showed that the traditional approach can become more competitive and gain more organizational acceptance by addressing the lack of domain knowledge, improving cooperation with the requesting team, and developing soft skills in the development team. Each scenario resulted in an increase in organizational acceptance by 17–25% compared to the initial state of the traditional strategy.

This research emphasizes the importance of considering different implementation strategies for business intelligence systems and highlights the potential benefits of utilizing a dynamic model to simulate and improve these strategies. By implementing the recommendations from this study, organizations can increase their chances of successful implementation and gain more support for their business intelligence systems. The system dynamics tool provides researchers with the possibility of modeling and simulating complex peripheral phenomena, leading to more logical and accurate decisions based on the results. Thus, this study creates a framework for future research in the area of business intelligence system implementation strategies.

**Limitations of the study**

There are a few limitations to this study that should be acknowledged. First, the research is based on interviews with only five BI experts. While their opinions and insights were valuable, they may not represent the full range of perspectives on BI implementation strategies. Future research could expand the sample size to include a broader range of stakeholders, including IT managers, business analysts, and end-users.

Second, the simulation model used in this study is based on certain assumptions about the behavior of users and the organization over time. The accuracy of the model depends on the validity of these assumptions, which may not always hold in real-world settings. Future research could use alternative modeling approaches or validate the assumptions through field studies or experiments.

Third, this study focuses exclusively on organizational acceptance as an outcome variable. Other important outcomes of BI implementation, such as user satisfaction, system performance, and business impact, were not considered. Future research could examine these additional outcomes and explore the relationships between them and organizational acceptance.

**Future research directions**

This study provides a foundation for future research in the area of BI implementation strategies. One potential direction for future research is to investigate the role of data quality and data governance in BI adoption. Data quality and governance are critical factors that can directly affect the usefulness and reliability of BI systems. Understanding how these factors interact with implementation strategies and organizational acceptance can provide valuable insights for improving BI adoption.

Another direction for future research is to explore the impact of organizational culture on BI implementation success. Organizational culture can shape attitudes and behaviors toward BI adoption, and understanding its role can help organizations develop more effective implementation strategies. Future research could examine the relationship between organizational culture, implementation strategies, and organizational acceptance.

Finally, future research could further explore the potential benefits of using system dynamics modeling in the context of BI implementation. System dynamics can enable researchers to capture the complexity of organizational processes and interactions and simulate different scenarios to test and refine implementation strategies. Further research could investigate the effectiveness of system dynamics modeling in other contexts and explore ways to enhance its accuracy and validity.

**Declaration of Competing Interest**

The authors declare that they have no known competing financial interests or personal relationships that could have appeared to influence the work reported in this paper.

**Supplementary materials**

Supplementary material associated with this article can be found, in the online version, at doi:10.1016/j.teler.2023.100070.

**References**

[1] N. Ain, G. Vaia, W.H. DeLone, M. Waheed, Two decades of research on business intelligence system adoption, utilization and success–a systematic literature review, Decis. Support Syst. 125 (2019) 113113.
[2] P. Alpar, M. Schulz, Self-service business intelligence, Bus. Inf. Syst. Eng. 58 (2) (2016) 151–155.
[3] A.L. Antunes, E. Cardoso, J. Barateiro, Incorporation of ontologies in data warehouse/business intelligence systems-a systematic literature review, Int. J. Inf. Manag. Data Insights 2 (2) (2022) 100131.
[4] H. Ashok, S. Ayyasamy, A. Ashok, V. Arunachalam, E-business analytics through ETL and self-service business intelligence tool, 2020S International Conference on Inventive Research in Computing Applications (ICIRCA), 2020.
[5] A. Audzeyeva, R. Hudson, How to get the most from a business intelligence application during the post implementation phase? Deep structure transformation at a UK retail bank, Eur. J. Inf. Syst. 25 (1) (2016) 29–46.
[6] M.P. Bach, A. Čeljo, J. Zoroja, Technology acceptance model for business intelligence systems: preliminary research, Procedia Comput. Sci. 100 (2016) 995–1001.
[7] E. Ballard, A. Farrell, M. Long, Community-Based system dynamics for mobilizing communities to advance school health, J. Sch. Health 90 (12) (2020) 964–975.
[8] M. Bastan, M. Zarei, R. Tavakkoli-Moghaddam, A new technology acceptance model: a mixed-method of grounded theory and system dynamics, Kybernetik 51 (1) (2022) 1–30.
[9] A. Blouchoutzi, G. Tsaples, D. Manou, J. Papathanasiou, Investigating public-private cooperation in migrant labor market integration: a system dynamics study to explore the challenge for Greece, Economies 11 (2) (2023) 38 https://www.mdpi.com/2227-7099/11/2/38.
[10] C. Brewis, S. Dibb, M. Meadows, Leveraging Big Data for Strategic Marketing: a dynamic capabilities model for incumbent firms, Technol. Forecast. Soc. Change 190 (2023) 122402.
[11] Y. Cao, Z.J. Zhou, C.H. Hu, S.W. Tang, J. Wang, A new approximate belief rule base expert system for complex system modelling, Decis. Support Syst 150 (2021) 113558.
[12] N. Caseiro, A. Coelho, Business intelligence and competitiveness: the mediating role of entrepreneurial orientation, Compet. Rev. 28 (2) (2018) 213–226.
[13] M. David Stone, N. David Woodcock, Interactive, direct and digital marketing: a future that depends on better use of business intelligence, J. Res. Interact. Market. 8 (1) (2014) 4–17.
[14] N.A. El-Adaileh, S. Foster, Successful business intelligence implementation: a systematic literature review, J. Work-Appl. Manag. (2019).
[15] L.Y. Fang, N.F.M. Azmi, Y. Yahya, H. Sarkan, N.N.A. Sjarif, S. Chuprat, Mobile business intelligence acceptance model for organisational decision making, Bull. Electric. Eng. Inf. 7 (4) (2018) 650–656.
[16] J. Fjermestad, S. Kudyba, K. Lawrence, in: Business Intelligence and Analytics Case Studies, Taylor & Francis, 2018, pp. 77–78. Vol. 28.
[17] D.N. Ford, A system dynamics glossary, Syst. Dyn. Rev. 35 (4) (2019) 369–379.
[18] H.-P. Fu, T.-H. Chang, Y.-H. Teng, C.-H. Liu, H.-C. Chuang, Critical factors considered by companies to introduce business intelligence systems, Axioms 11 (7) (2022) 338.
[19] R. Gaardboe, T.S. Jonasen, Business intelligence success factors: a literature review, J. Inf. Technol. Manag. 29 (1) (2018) 1–15.
[20] S. Goundar, K. Okafor, A. Cagica, P. Chand, S. Singh, Using business intelligence in organizations, in: Enterprise Systems and Technological Convergence: Research and Practice, 2021, p. 99.
[21] T. Grubljesič, P.S. Coelho, J. Jaklič, The shift to socio-organizational drivers of business intelligence and analytics acceptance, J. Organ. End User Comput. 31 (2) (2019) 37–64.
[22] T. Grubljesič, J. Jaklič, Business intelligence acceptance: the prominence of organizational factors, Inf. Syst. Manag. 32 (4) (2015) 299–315.
[23] R. Harrison, A. Parker, G. Brosas, R. Chiong, X. Tian, The role of technology in the management and exploitation of internal business intelligence, J. Syst. Inf. Technol. (2015).
[24] L. Harst, H. Lantzsch, M. Scheibe, Theories predicting end-user acceptance of telemedicine use: systematic review, J. Med. Internet Res. 21 (5) (2019) e13117, doi:10.2196/13117.
[25] J. Hou, L. Wang, Z. Lin, M. Gao, Research on operational effectiveness evaluation of network information system based on system dynamics, 2021 International Conference on Electronics, Circuits and Information Engineering (ECIE), 2021.
[26] Z. Hussain, A. Jabbar, K. Kong, Power, Dominance and Control: Implementing a New Business Intelligence System, Digital Transformation and Society, 2023.
[27] T. Jafari, A. Zarei, A. Azar, A. Moghaddam, The impact of business intelligence on supply chain performance with emphasis on integration and agility–a mixed research approach, Int. J. Prod. Perform. Manag. (2021).
[28] M. Jalilvand Khosravi, M. Maghsoudi, S Salavatian, Identifying and clustering users of VOD platforms using SNA technique: a case study of cinemamarket, New Market. Res. J. 11 (4) (2022) 20–21.
[29] Z. Jamshidi, S.M. Sajadi, K. Talebi, S.H. Hosseini, Applying system dynamics approach to modelling growth engines in the international entrepreneurship era, in: Empirical International Entrepreneurship, Springer, 2021, pp. 491–513.
[30] M.A. Kermani, M. Maghsoudi, M.S. Hamedani, A. Bozorgipour, Analyzing the interorganizational collaborations in crisis management in coping with COVID-19 using social network analysis: case of Iran, J. Emerg. Manag. 20 (3) (2022) 249–266.
[31] A. Kumar, B. Krishnamoorthy, Business analytics adoption in firms: a qualitative study elaborating TOE framework in India, Int. J. Glob. Bus. Compet. 15 (2) (2020) 80–93.
[32] C. Lennerholt, Facilitating the Implementation and Use of Self Service Business Intelligence, University of Skövde, 2022 ].
[33] C. Lennerholt, J.v. Laere, E. Söderström, Success factors for managing the SSBI challenges of the AQUIRE framework, J. Decis. Syst. (2022) 1–22, doi:10.1080/12460125.2022.2057006.
[34] C. Lennerholt, J. van Laere, Data access and data quality challenges of self-service business intelligence, 27th European Conference on Information Systems (ECIS), 2019.
[35] C. Lennerholt, J. van Laere, E. Söderström, Implementation challenges of self service business intelligence: a literature review, 51st Hawaii International Conference on System Sciences, 2018.
[36] C. Lennerholt, J. Van Laere, E. Söderström, User-related challenges of self-service business intelligence, Inf. Syst. Manag. 38 (4) (2021) 309–323, doi:10.1080/10580530.2020.1814458.
[37] P.M. Leonardi, COVID-19 and the new technologies of organizing: digital exhaust, digital footprints, and artificial intelligence in the wake of remote work, J. Manag. Stud. 58 (1) (2021) 249.
[38] Q. Li, L. Zhang, L. Zhang, S. Jha, Exploring multi-level motivations towards green design practices: a system dynamics approach, Sustain. Cities Soc. 64 (2021) 102490.
[39] X. Li, J.P.-A. Hsieh, A. Rai, Motivational differences across post-acceptance information system usage behaviors: an investigation in the business intelligence systems context, Inf. Syst. Res. 24 (3) (2013) 659–682.
[40] M.R. Llave, Data lakes in business intelligence: reporting from the trenches, Procedia Comput. Sci. 138 (2018) 516–524.
[41] E. Malbon, J. Parkhurst, System dynamics modelling and the use of evidence to inform policymaking, Policy Stud. (2022) 1–19.
[42] M.I. Merhi, Evaluating the critical success factors of data intelligence implementation in the public sector using analytical hierarchy process, Technol. Forecast. Soc. Change 173 (2021) 121180.
[43] A. Mousavi, M. Mohammadzadeh, H. Zare, Developing a system dynamic model for product life cycle management of generic pharmaceutical products: its relation with open innovation, J. Open Innov. 8 (1) (2022) 14 https://www.mdpi.com/2199-8531/8/1/14.
[44] J. Müller, G. Schuh, D. Meichsner, G. Gudergan, Success factors for implementing Business Analytics in small and medium enterprises in the food industry, 2020 IEEE International Conference on Technology Management, Operations and Decisions (ICTMOD), 2020.
[45] S. Nalchigar, E. Yu, S. Easterbrook, Towards actionable business intelligence: can system dynamics help? IFIP Working Conference on The Practice of Enterprise Modeling, 2014.
[46] N. Nithya, R. Kiruthika, Impact of Business Intelligence Adoption on performance of banks: a conceptual framework, J. Ambient Intell. Humaniz Comput. 12 (2021) 3139–3150.
[47] Y. Niu, L. Ying, J. Yang, M. Bao, C. Sivaparthipan, Organizational business intelligence and decision making using big data analytics, Inf. Process. Manag. 58 (6) (2021) 102725.
[48] C.I. Papanagnou, O. Matthews-Amune, Coping with demand volatility in retail pharmacies with the aid of big data exploration, Comput. Oper. Res. 98 (2018) 343–354.
[49] J. Passlick, L. Grützner, M. Schulz, M.H. Breitner, Self-service business intelligence and analytics application scenarios: a taxonomy for differentiation, Inf. Syst. e-Bus. Manag. (2023), doi:10.1007/s10257-022-00574-3.
[50] M.D. Peters, B. Wieder, S.G. Sutton, J. Wakefield, Business intelligence systems use in performance measurement capabilities: implications for enhanced competitive advantage, Int. J. 21 (2016) 1–17.
[51] G. Phillips-Wren, M. Daly, F. Burstein, Reconciling business intelligence, analytics and decision support systems: more data, deeper insight, Decis. Support Syst. 146 (2021) 113560.
[52] I. Pluchinotta, A. Pagano, T. Vilcan, S. Ahilan, L. Kapetas, S. Maskrey, V. Krivtsov, C. Thorne, E O'Donnell, A participatory system dynamics model to investigate sus-

tainable urban water management in Ebbsfleet Garden City, Sustain. Cities Soc. 67 (2021) 102709.
[53] M.J. Radzicki, System dynamics and its contribution to economics and economic modeling, in: System Dynamics: Theory and Applications, 2020, pp. 401–415.
[54] J. Ranjan, C. Foropon, Big data analytics in building the competitive intelligence of organizations, Int. J. Inf. Manage. 56 (2021) 102231.
[55] A.Z. Ravasan, S.R. Savoji, An investigation of BI implementation critical success factors in Iranian context, Int. J. Bus. Intell. Res. 5 (3) (2014) 41–57, doi:10.4018/ijbir.2014070104.
[56] F. Ricciardi, P. De Bernardi, V. Cantino, System dynamics modeling as a circular process: the smart commons approach to impact management, Technol. Forecast. Soc. Change 151 (2020) 119799.
[57] K. Saeed, A. Sidorova, A. Vasanthan, The bundling of business intelligence and analytics, Int. J. Comput., Inf., Syst. Sci., Eng. (2022) 1–12.
[58] G.B. Sajons, Estimating the causal effect of measured endogenous variables: a tutorial on experimentally randomized instrumental variables, Leadersh Q 31 (5) (2020) 101348.
[59] D. Schuff, K. Corral, R.D. St. Louis, G Schymik, Enabling self-service BI: a methodology and a case study for a model management warehouse, Inf. Syst. Front. 20 (2018) 275–288.
[60] S. Shafiee, S. Jahanyan, A.R. Ghatari, A. Hasanzadeh, Developing sustainable tourism destinations through smart technologies: a system dynamics approach, J. Simul. (2022) 1–22.
[61] F. Sönmez, Technology acceptance of business intelligence and customer relationship management systems within institutions operating in capital markets, Int. J. Acad. Res. Bus. Soc. Sci. 8 (2) (2018) 400–422.
[62] V. Srikrishnan, D.C. Lafferty, T.E. Wong, J.R. Lamontagne, J.D. Quinn, S. Sharma, N.J. Molla, J.D. Herman, R.L. Sriver, J.F. Morris, Uncertainty analysis in multi-sector systems: considerations for risk analysis, projection, and planning for complex systems, Earth's Fut. 10 (8) (2022) e2021EF002644.
[63] C.A. Tavera Romero, J.H. Ortiz, O.I. Khalaf, A. Ríos Prado, Business intelligence: business evolution after industry 4.0, Sustainability 13 (18) (2021) 10026.
[64] J.R. Turner, R.M. Baker, Complexity theory: an overview with potential applications for the social sciences, Systems 7 (1) (2019) 4.
[65] N. Ul-Ain, G. Vaia, W. DeLone, Business intelligence system adoption, utilization and success-A systematic literature review, in: Proceedings of the 52nd Hawaii International Conference on System Sciences, 2019.
[66] P.D. Vecchio, G. Secundo, Y. Maruccia, G. Passiante, A system dynamic approach for the smart mobility of people: implications in the age of big data, Technol. Forecast. Soc. Change 149 (2019) 119771, doi:10.1016/j.techfore.2019.119771.
[67] N. Vella, Business Intelligence and Data-Driven Decision-Making: A Management Accounting Perspective, University of Malta, 2021.
[68] M.A. Villanthenkodath, M.A. Ansari, P. Kumar, Y.N. Raju, Effect of information and communication technology on the environmental sustainability: an empirical assessment for South Africa, Telemat. Inform. 7 (2022) 100013, doi:10.1016/j.teler.2022.100013.
[69] L.L. Visinescu, M.C. Jones, A. Sidorova, Improving decision quality: the role of business intelligence, Int. J. Comput., Inf., Syst. Sci., Eng. 57 (1) (2017) 58–66.
[70] M. Wee, Business Intelligence & Analytics Adoption in Australian SMEs: Identified Processes, Decision-Making, and Leadership Skills, Swinburne University of Technology, 2021 ].
[71] P. Weichbroth, J. Kowal, M. Kalinowski, Toward a unified model of mobile Business Intelligence (m-BI) acceptance and use, in: Proceedings of the 55th Hawaii International Conference on System Sciences, 2022.
[72] W. Yeoh, A. Popovič, Extending the understanding of critical success factors for implementing business intelligence systems, J. Assoc. Inf. Sci. Technol. 67 (1) (2016) 134–147.
[73] F. Zare, S. Elsawah, A. Bagheri, E. Nabavi, A.J. Jakeman, Improved integrated water resource modelling by combining DPSIR and system dynamics conceptual modelling techniques, J. Environ. Manage. 246 (2019) 27–41.
[74] M. Zhang, M. Cheng, Big data, social media, and intelligent communication, Telemat. Inform. 8 (2022) 100026, doi:10.1016/j.teler.2022.100026.
[75] X. Zhang, H. Ding, Y. Wang, Study on business intelligence products supporting knowledge management, in: Frontiers in Enterprise Integration, CRC Press, 2020, pp. 213–222.